\documentclass[10pt]{article}
\usepackage{spconf,amsmath,graphicx,multirow}
\usepackage[colorlinks,linkcolor=red]{hyperref}
\usepackage{booktabs}
\usepackage{anyfontsize}
\usepackage{amssymb}
\usepackage{enumitem}
\usepackage{caption}
\usepackage{hyperref}

\ninept
\title{Community detection graph convolutional network for Overlap-aware Speaker Diarization}
\name{Jie Wang$^{\dag}$$^{1}$, Zhicong Chen$^{\dag}$$^{1}$, Haodong Zhou$^{1}$, Lin Li$^{*1}$, Qingyang Hong$^{2}$
\thanks{$^{\dag}$ Equally contributed.\quad $^{*}$ Corresponding author.}
\thanks{
This work was supported in part by the National Natural Science Foundation of China under Grants 62001405, 62276220, and 61876160.
}}
\address{ $^1$School of Electronic Science and Engineering, Xiamen University, China\\
  $^2$School of Informatics, Xiamen University, China\\
  lilin@xmu.edu.cn}

\begin{document}

%
\maketitle
\begin{abstract}
The clustering algorithm plays a crucial role in speaker diarization systems. However, traditional clustering algorithms suffer from the complex distribution of speaker embeddings and lack of digging potential relationships between speakers in a session. We propose a novel graph-based clustering approach called Community Detection Graph Convolutional Network (CDGCN) to improve the performance of the speaker diarization system. The CDGCN-based clustering method consists of graph generation, sub-graph detection, and Graph-based Overlapped Speech Detection (Graph-OSD). Firstly, the graph generation refines the local linkages among speech segments. Secondly the sub-graph detection finds the optimal global partition of the speaker graph. Finally, we view speaker clustering for overlap-aware speaker diarization as an overlapped community detection task and design a Graph-OSD component to output overlap-aware labels. By capturing local and global information, the speaker diarization system with CDGCN clustering outperforms the traditional Clustering-based Speaker Diarization (CSD) systems on the DIHARD III corpus.
\end{abstract}
\section{Introduction}
\label{sec:introduction}
Speaker diarization is a problem of grouping speech segments in
an audio recording according to the speakers’ identities. We have witnessed the rising popularity of speaker diarization over recent years for its significant applications of minutes of meetings, multi-speaker transcription, pre-processing for automatic speech recognition (ASR) \cite{boeddeker2018front}\cite{medennikov2020stc}, and so on. As the deployments for scenarios have grown in complexity, speaker diarization systems confront many difficulties, such as the unknown number of speakers and handling the overlapped speech.

\begin{figure}[htb]
\centering
\centerline{\includegraphics[width=8cm, trim=0 0 420 150,clip]{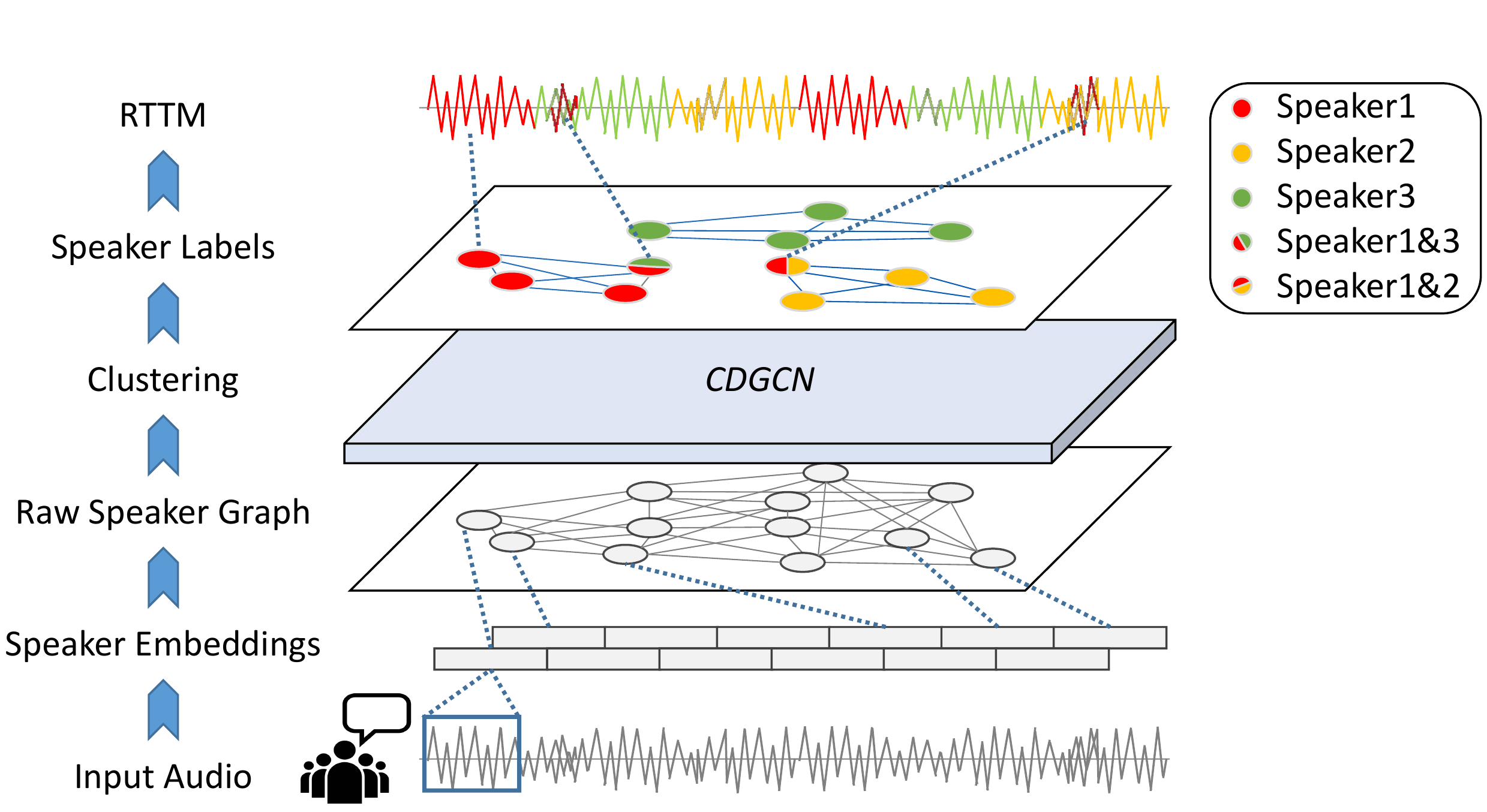}}

\caption{An illustration of the speaker diarization system pipeline with the CDGCN clustering method. The Rich Transcription Time Marked (RTTM) is the output of the speaker diarization systems.}
\label{fig:pipeline}
\vspace{-0.5cm}
\end{figure}

Clustering-based approaches are widely used in speaker diarization because it allows for flexible and scalable speaker modeling using various techniques \cite{shum2013unsupervised}\cite{sell2014speaker}\cite{ning2006spectral}. There are three modules in clustering-based Speaker Diarization (CSD) systems: speaker embedding extractor, clustering module, and post-processing module. Typically, the clustering modules in CSD systems utilize conventional clustering algorithms, such as Agglomerative Hierarchical Clustering (AHC) \cite{sell2016priors}\cite{xiao2021microsoft}, Spectral Clustering (SC) \cite{wang22ha_interspeech}\cite{lin2020self}\cite{lin19_interspeech} and K-means\cite{wang2018speaker}, to perform speaker embeddings clustering. However, traditional conventional clustering algorithms suffer from complicated distribution of speaker embeddings \cite{shi2000normalized} and is sensitive to hyper-parameter. For example, SC assumes the sizes of clusters are relatively balanced, while K-means assumes the clusters are spherical. Moreover, the performance of AHC is affected by the threshold sensitively. These assumptions limit the speaker clustering performance and degrade the diarization quality.

The distribution of speakers is hard to be modeled with Euclidean structures, because of the complex interrelation among speakers. Graph Convolutional Network (GCN) \cite{welling2016semi} is proposed to handle the data of non-Euclidean structure. Many GCN-based clustering methods are recently proposed for large-scale embeddings clustering instead of relying on hand-crafted criteria. Tong et al. \cite{tong2022graph} adopted Detection Segmentation Graph Convolutional Network (DSGCN) for semi-supervised speaker recognition. Wang et al. \cite{wang2020speaker} used a GCN to refine speaker embeddings for affinity matrix on speaker diarization system.

Inspired by these works, we proposed a new GCN-based clustering approach with community detection for speaker diarization named Community Detection Graph Convolutional Network (CDGCN). We regard the clustering of speaker embeddings as a speaker graph generation and sub-graph detection task. The key idea is to build a refined speaker graph for segment embeddings and globally partition speaker graph to assign speaker labels for segments. The CDGCN-based clustering method also can assign multi-labels for each node to handle overlapped speech. 

The remainder of this paper is organized as follows. In Section \ref{sec:Review}, we revisit graph convolutional networks. The proposed approach is addressed in Section \ref{sec:approach}. In Section \ref{sec:data}, we describe the dataset of our experiments. In Section \ref{sec:experiment}, we evaluate the proposed systems on the DIHARD III \cite{ryant2020third}. Finally Section \ref{sec:conclusions} concludes this work.

\section{graph convolutional network}
\label{sec:Review}
In our work, a modified GCN \cite{wang2019linkage} model was adopted to build speaker graphs. The input of the GCN model is an embedding matrix $\boldsymbol{H} \in \mathbb{R}^{K \times D}$ together with an adjacency matrix $\boldsymbol{A} \in \mathbb{R}^{K \times K}$, where ${K}$ is the number of nodes in a graph and ${D}$ is the dimension of the embeddings. The feedforward of the GCN model can be summarized in two steps:

(1) \textbf{Aggregation}: The aggregation processing allows each node to learn the information from neighbors on the graph. After graph aggregation, the GCN layer transforms $\boldsymbol{H}^{(l)}$ into a hidden feature matrix $\boldsymbol{H}^{(l+1)}$. The aggregation is formulated as follows:
\vspace{-0.8cm}
\begin{center}
\begin{equation}
\boldsymbol{H}^{(l+1)}=\sigma([\boldsymbol{H}^{(l)}\parallel \boldsymbol{\hat{A}}\boldsymbol{H}^{(l)}]\boldsymbol{W}^{(l)})
\end{equation}
\end{center}
\vspace{-0.2cm}
where $\boldsymbol{H}^{(l)} \in \mathbb{R}^{K \times D^{(l)}}$, $\boldsymbol{H}^{(l+1)} \in \mathbb{R}^{K \times D^{(l+1)}}$ denotes the output data with ${D^{(l+1)}}$ dimensions in ($l$+1)-th layer, $\sigma$ is the Relu activation function, $\boldsymbol{W}^{(l)} \in \mathbb{R}^{2D^{(l)} \times D^{(l+1)}}$ is a learnable weight matrix in the $l$-th layer, $\boldsymbol{\hat{A}}$ is the normalized and regularized affinity matrix with $K \times K$ size and each row is summed up to 1. ``$\parallel$" denotes matrix concatenation operation along the feature dimension. The normalized affinity matrix $\boldsymbol{\hat{A}}$ is formulated as:

\vspace{-0.8cm}
\begin{center}
\begin{equation}
\boldsymbol{\hat{A}}=\boldsymbol{\widetilde{D}}^{-\frac{1}{2}} \boldsymbol{\widetilde{A}} \boldsymbol{\widetilde{D}}^{-\frac{1}{2}}
\end{equation}
\end{center}
\vspace{-0.2cm}

where, $\boldsymbol{\widetilde{A}}=\boldsymbol{A}+\boldsymbol{I}$ is the adjacency matrix with self connection, $\boldsymbol{I}$ is the unit matrix and $\boldsymbol{\widetilde{D}}$ denotes the degree matrix of $\boldsymbol{\widetilde{A}}$ with $\boldsymbol{\widetilde{D}}_{i i}=\sum_j \boldsymbol{\widetilde{A}}_{i j}$. 

(2) \textbf{Prediction}: Finally, the prediction labels of nodes $\boldsymbol{Y}=\{y_1, y_2,...,y_K\} \in \mathbb{R}^{K}$ are generated by two stacked linear layers with a softmax function. The labeling principle is that $y_k$=1 if there is a linkage between the pivot node and the $k$-th node; otherwise reverse. The GCN is trained by Binary Cross Entropy (BCE) loss.

\section{Proposed approach}
\label{sec:approach}

\subsection{System pipeline}
\label{subsec:pipeline}
The CDGCN-based speaker diarization system pipeline is shown in Figure \ref{fig:pipeline}. Firstly, input samples are split into segments with slide windows. Then, the embedding extractor converts speech segments into fixed dimension vectors called x-vectors $\boldsymbol{X} \in \mathbb{R}^{N \times D}$ where $N$ is the number of segments and $D$ is the feature dimension of an embedding. We adopt a ResNet-34-SE model to build the extractor. After that, we construct the raw speaker graph by calculating cosine similarity scores between embeddings. The CDGCN-based clustering module takes the raw speaker graph and outputs overlap-aware speaker labels. The diarization results follow from the labels.

\begin{figure*}[htb]
  \centering
  \centerline{\includegraphics[width=16.5cm, trim=0 60 0 105,clip]{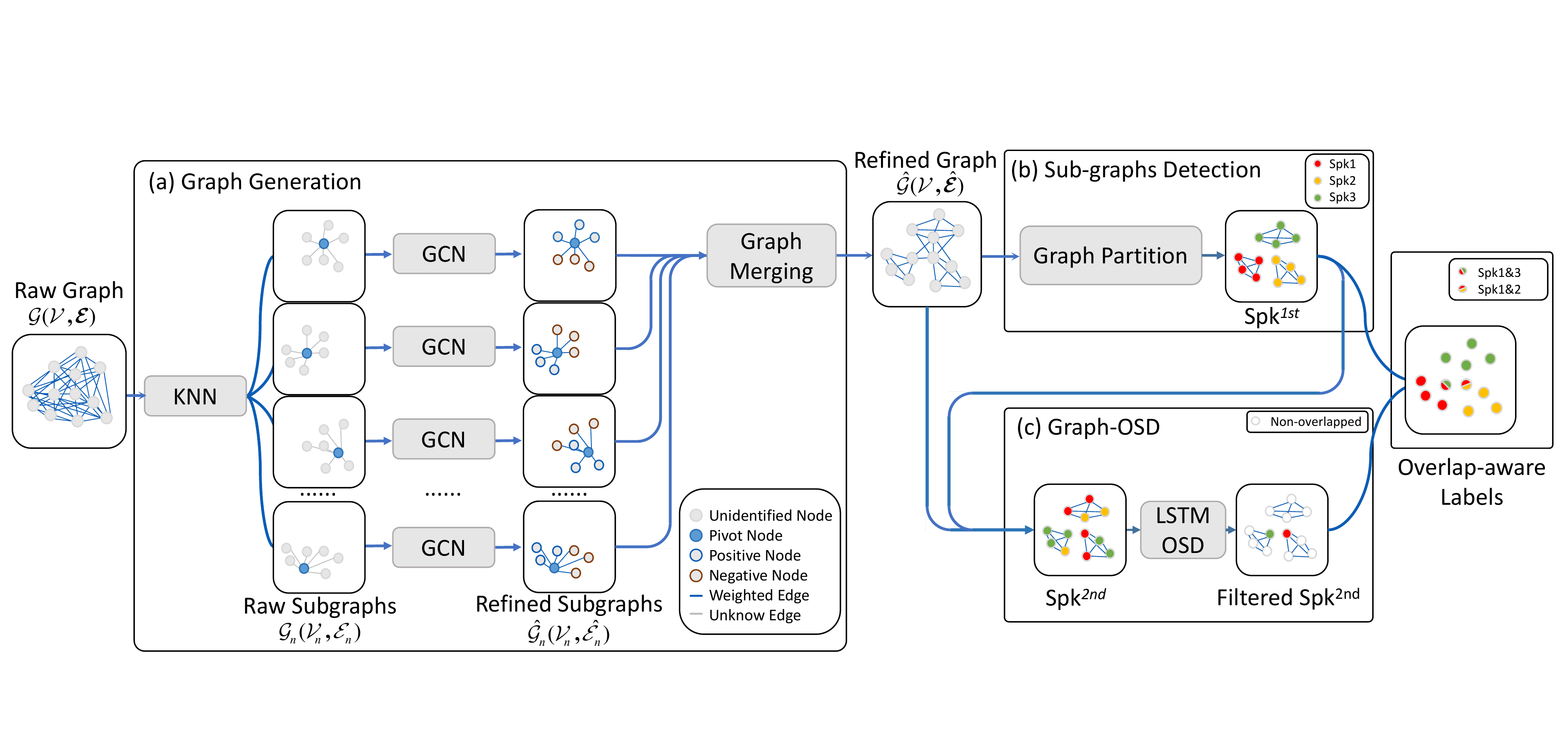}}

\caption{The architecture of Community Detection Graph Convolutional Network based clustering algorithm. ``Spk$^{1st}$" denotes the most-likely speaker labels of nodes, and ``Spk$^{2nd}$" indicates the second most-likely speaker labels of nodes.}
\label{fig:cdgcn}
\vspace{-0.6cm}
\end{figure*}

\subsection{CDGCN-based clustering}
The overall block diagram of CDGCN is shown in Figure \ref{fig:cdgcn}. The basic concept of the CDGCN is to estimate the topological connection of speech segments and use a community detection algorithm to find the optimal partitions. CDGCN-based clustering method contains graph generation, sub-graphs detection, and graph-based overlapped speech detection (Graph-OSD). Each component of CDGCN will be described as follows.

\subsubsection{Graph Generation}
The input of clustering module is a raw graph ${\mathcal{G}}=(\mathcal{V},\boldsymbol{\mathcal{E})}$, where nodes $\mathcal{V}=\{{v}_{1}, {v}_{2},...,{v}_{N}\} \in \mathbb{R}^{N}$ represent speech segments, edges $\boldsymbol{\mathcal{E}} = \{\boldsymbol{e}_{1},\boldsymbol{e}_{2},...,\boldsymbol{e}_{N}\} \in \mathbb{R}^{N \times N}$ are cosine similarity scores between pairs of embeddings and $N$ is the number of segments. The raw graph is a complex full-connected graph which is vulnerable to noise. In order to tackle this problem, we design a graph generation to refine interrelations between speech segments according to local context information. Firstly, we adopt the $K$-Nearest Neighbors (KNN) algorithm to create sub-graphs for each node. The sub-graph $\mathcal{G}_{n}=(\mathcal{V}_{n},\mathcal{E}_{n})$ is built for $n$-th pivot node, where $\mathcal{V}_{n} \in \mathbb{R}^{K}$ is the top-$K$ nearest neighbor of pivot node and $\mathcal{E}_{n} \in \mathbb{R}^{K}$ denotes the similarity among $n$-th pivot node and its neighbors. For example, as shown in Figure \ref{fig:cdgcn} (a), let $K$ and the nodes number $N$ be 6 and 12 respectively. The raw speaker sub-graphs $\mathcal{G}_{n}$ are fed into GCN model mentioned in Section \ref{sec:Review} and the refined sub-graphs $\mathcal{\hat{G}}_{n}=(\mathcal{V}_{n},\mathcal{\hat{E}}_{n})$ are predicted, where $\mathcal{\hat{E}}_{n}=\{\hat{e}_{n}^{1},\hat{e}_{n}^{2},...,\hat{e}_{n}^{K}\}\in \mathbb{R}^{K}$ are predicted edges and $\hat{e}_{n}^{K}$ indicates the probability that pivot node and $k$-th node belong to the same cluster. Then, the refined speaker sub-graphs are merged to acquire the total refined 
speaker graph $\mathcal{\hat{G}}=(\mathcal{V},\mathcal{\hat{E}})$ which is a weighted undirected graph. In the graph merging stage, multiple edges between two nodes keep the bigger one.

\subsubsection{Sub-graphs Detection}
One of the main obstacles is to partition the refined graph robustly. The sub-graphs detection predicts the most-likely community label of nodes on the speaker graph. Zheng et al. \cite{zheng2022reformulating} use Leiden community detection\cite{V2019From} with Uniform Manifold Approximation and Projection (UMAP) for speaker clustering on the simulated meetings. In this part, we adopt Leiden community detection for sub-graphs detection. Community is interpreted as clusters of densely interconnected nodes that are only sparsely connected with the rest on the graph \cite{PhysRevE.74.016110}. In our work, each community label corresponds to a speaker label. Community detection aims to group nodes with an optimization quality function. The higher the Q is, the better the clustering result we may obtain. The quality function $Q$ \cite{V2019From} is represented as: 
\vspace{-0.8cm}
\begin{center}
\begin{equation}
Q = \sum_{c} \left(m_c - \gamma\frac{K_c^2}{4m} \right)
\end{equation}
\end{center}
\vspace{-0.1cm}
where $m_c$ is total internal edge weight of community $c$, $m$ is the total number of edges, $K_c$ is the total weighted degree of nodes in community $c$, and $\gamma$ is a resolution parameter that controls the number of communities. $K_c$ is given by
\vspace{-0.8cm}
\begin{center}
\begin{equation}
K_c = \sum_{i \mid \sigma_i = c} k_i
\end{equation}
\end{center}
\vspace{-0.1cm}
here, $\sigma_i$ denotes the community label of node $i$, and $k_i$ is the weighted degree of node $i$. 

The Leiden community detection consists of the following phases:

(1) \textbf{Initial partition}: The Leiden algorithm assigns each node to a singleton community.

(2) \textbf{Nodes Local moving}: The individual node is moved from one community to another to find a better partition $P$ with higher Q. 

(3) \textbf{Partition refinement}: In the refinement phase, the refined partition $P_{refined}$ is initially set to a singleton partition. And then, the nodes in each community are merged locally to refine partition $P_{refined}$. After performing refinement, communities in $P$ may be split into subcommunities.

(4) \textbf{Graph aggregation}: An aggregation graph is constructed based on $P_{refined}$. In this phase, the node belonging to the same community are merged into a new node. 

(5) \textbf{Iteration}: Phases 2-4 are repeated until no further improvements of quality function can be made.

\subsubsection{Graph-OSD}

Overlapped speech handling is the critical processing of speaker diarization. In this work, we propose a Graph-based Overlapped Speech Detection (Graph-OSD) module in CDGCN algorithm. As shown in Figure \ref{fig:cdgcn} (c), we view the speaker clustering of diarization as overlapped community detection task. The Graph-OSD is a two-stage model to handle overlapped speech and assume there are at most two speakers at once. 

In the first stage, we predict the second community label for each node. According to the refined graph, and the partition created by sub-graphs detection, we calculate the belonging coefficient $b_{(c,i)}$ for each node $i$, where $b_{(c,i)}$ presents the strength of membership that $i$-th node belongs to community $c$. This process is defined as:
\vspace{-0.5cm}
\begin{center}
\begin{equation}
b_{(c,i)} = \sum_{j \mid \sigma_j = c} e_{i j}
\end{equation}
\end{center}
here, $\sigma_j$ denotes the community label of node $j$, and $e_{i j}$ is the weighted edge between node $i$ and node $j$ from refined graph. Based on the most-likely community label and belonging coefficient, the second most-likely community $\widetilde{c}_i$ of node $i$ is given by
\begin{equation}
\tilde{c}_{i}=\underset{c \in C,c \neq \hat{c}_{i}}{\arg\max}\, b_{(c, i)}
\end{equation}
where $C$ is the estimated communities number and $\hat{c}_i$ is the most-likely community label.

The next stage predicts the overlapped speech regions and ignores the second speaker labels at non-overlapped regions. We perform an LSTM-based OSD model described in \cite{Bredin2021} to predict the frame-level overlapped/non-overlapped regions of speech. The model is an end-to-end overlapped speech detection whose output is a frame-level binary sequence, and is trained with the binary cross entropy loss function. Finally, we output the two most likely speakers for each frame in overlapped speech region.

\section{Datasets and experimental setup}
\subsection{Data preparation}
We evaluate our speaker diarization systems on the DIHARD III corpus. The DIHARD III contains the development (DEV) and evaluation (EVAL) set from 11 domains exhibiting wide variation in equipment. The overlap ratio of DIHARDIII Core and Full dataset is 8.75\% and 9.35\%, respectively. The detailed training sets of different modules on our speaker diarization systems are described as follows.
\label{sec:data}
\begin{itemize}[leftmargin=*]
\item Speaker embedding extractor: We train the embedding extractor with the VoxCeleb2 dataset. The VoxCeleb2 contains over 1 million utterances from 5,994 speakers. 
\item GCN: We extracted 256-dimensional embeddings for VoxCeleb2. We constructed the sub-graph for each utterance. Each sub-graph is a training instance of the GCN model.

\item LSTM-based OSD: We adopted the DIHARD III DEV to train the OSD module. 
\end{itemize}
\subsection{Experimental setup}
 During training, we extracted the 81-dimensional log-mel filter-bank (FBank) with a window size of 25ms and a 10ms shift. In our diarization systems, we split the audio into 1.5s length segments with 0.75s windows shift, and extracted the embeddings of segments with the ResNet-34-SE model from ASV-Subtools\cite{tong2021asv}. In the GCN module, we stacked four GCN layers and set the $K$ of KNN to 300. The resolution $\gamma$ of Leiden community detection module is set to 0.6 and the threshold of AHC is set to 0.17.

\begin{table*}[]
\centering
\tabcolsep=0.5cm
\caption{The comparison among different CSD systems on DIHARD III with 0ms collar condition. We evaluated the diarization systems are evaluated on core and full datasets with oracle Voice Activity Detection (VAD). The core is a subset of the full evaluation set and strives for balance cross-domains. DOVER-Lap \cite{raj2021dover} is a subsystems fusion algorithm.}
\label{tab:systems}
\begin{tabular}{clcccc}
\hline
                                            & \multicolumn{1}{c}{}                                   & \multicolumn{4}{c}{DER(\%)}                                                                                   \\ \cline{3-6} 
ID                                   & \multicolumn{1}{c}{Methods}                            & \multicolumn{2}{c}{DEV}                               & \multicolumn{2}{c}{EVAL}                              \\ \cline{3-6} 
                                            & \multicolumn{1}{c}{}                                   & Core                   & Full                      & Core                      & Full                      \\ \hline
\multirow{2}{*}{Official Baseline\cite{ryant2020third}} & TDNN+AHC                                               & 21.05                     & 20.71                     & 21.66                     & 20.75                     \\
                                            & TDNN+AHC+VB                                            & 20.25                     & 19.41                     & 20.65                     & 19.25                     \\ \hline
\multirow{4}{*}{Recent Works}               & ResNet+SC\cite{landini2021but}                              & \multicolumn{1}{l}{16.63} & \multicolumn{1}{l}{16.51} & \multicolumn{1}{l}{16.56} & \multicolumn{1}{l}{15.79} \\
                                            & ResNet+VBx\cite{landini2021but}                             & \multicolumn{1}{l}{16.66} & \multicolumn{1}{l}{16.26} & \multicolumn{1}{l}{16.67} & \multicolumn{1}{l}{15.74} \\
                                            & TDNN+VBx w/ OSD\cite{horiguchi2021hitachi}    & \multicolumn{1}{l}{\textbf{14.88}} & \multicolumn{1}{l}{13.87} & \multicolumn{1}{l}{18.20} & \multicolumn{1}{l}{15.65} \\
                                            & Res2Net+VBx w/ OSD (DOVER-Lap)\cite{horiguchi2021hitachi} & \multicolumn{1}{l}{15.18} & \multicolumn{1}{l}{14.04} & \multicolumn{1}{l}{18.47} & \multicolumn{1}{l}{15.81} \\ \hline
S1                                          & ResNet+AHC                                                    & 19.31                     & 19.94                     & 19.27                     & 18.90                     \\
S2                                          & ResNet+K-means                                                & \multicolumn{1}{l}{25.34} & \multicolumn{1}{l}{23.05} & \multicolumn{1}{l}{23.71} & \multicolumn{1}{l}{21.24} \\
S3                                          & ResNet+NME-SC                                                 & 18.56                     & 17.89                     & 17.98                     & 16.81                     \\
S4                                          & ResNet+CDGCN w/o Graph-OSD(ours)                                  & 17.10                     & 16.43                     & 16.50                     & 15.38                     \\
S5                                          & ResNet+CDGCN(ours)                                   & 15.40            & \textbf{13.67}            & \textbf{15.97}            & \textbf{13.72}            \\ \hline
\end{tabular}
\vspace{-0.3cm}
\end{table*}

\section{Experimental results}
\label{sec:experiment}

\subsection{Speaker clustering methods}
The first experiment explores the performance of different clustering algorithms on speaker diarization systems. The official baseline system provided by DIHARD III \cite{ryant2020third} consists of Time Delay Neural Network (TDNN) based x-vector extractor, Agglomerative Hierarchical Clustering (AHC) module, and Variational Bayes hidden Markov (VB) re-segmentation module. Our system pipeline is mentioned in Section \ref{subsec:pipeline}, and the difference among S1$\sim$S5 is the clustering method. For Systems S1$\sim$S3, we respectively performed AHC, K-means, and NME-SC (Normalized Maximum Eigengap Spectral Clustering)\cite{park2019auto} as clustering methods. In particular, we adopted Normalized Maximum Eigengap (NME)\cite{park2019auto} method to estimate the number of speakers for K-means. We performed the Leiden community detection algorithm on the CDGCN clustering method for system S4 and system S5. In order to investigate the effectiveness of the Graph-OSD module, we removed the module on system S4. In those systems, we tuned the hyper-parameters, including the threshold of AHC and resolution of CDGCN on DIHARD III DEV. 

The experimental results are shown in Table \ref{tab:systems}. We evaluated our systems under the same conditions as recent works \cite{ryant2020third}\cite{landini2021but}\cite{horiguchi2021hitachi}. By comparing the systems S1$\sim$S4, the experimental results showed that CDGCN assigned most-likely speaker labels for segments more accurately than other clustering algorithms. The results from system S5 demonstrated that the Graph-OSD module achieved better handling of overlapped speech.

\begin{table}[]
\centering

\caption{Ablation study on CDGCN-based speaker diarization system. + here denotes stacking our components of CDGCN. Oracle OSD indicates that the Graph-OSD replaces the overlapped speech label predicted by the LSTM model with ground truth labels.}
\label{tab:ablation}
\begin{tabular}{llcccc}
\hline
   & \multicolumn{1}{c}{}       & \multicolumn{4}{c}{DER(\%)}                                       \\ \cline{3-6} 
ID & \multicolumn{1}{c}{Method} & \multicolumn{2}{c}{DEV}         & \multicolumn{2}{c}{EVAL}        \\ \cline{3-6} 
   & \multicolumn{1}{c}{}       & Core           & Full           & Core           & Full           \\ \hline
S6 & Raw-Leiden                     & 24.92          & 22.03          & 25.18          & 21.59          \\
S7 & +KNN Graph                 & 18.57          & 17.70          & 18.58          & 17.04          \\
S4 & ++GCN refinement           & 17.10          & 16.43          & 16.50          & 15.38          \\
S5 & +++Graph-OSD         & \textbf{15.40} & \textbf{13.67} & \textbf{15.97} & \textbf{13.72} \\ \hline
S8 & ++++Oracle OSD             & 11.09          & 8.94           & 11.48          & 8.94           \\ \hline
\end{tabular}
\vspace{-0.5cm}
\end{table}

\vspace{-0.2cm}
\subsection{Ablation experiment}
We designed the second experiment to investigate the contribution of each module to CDGCN. As shown in Table \ref{tab:ablation}, we analyzed the gain from the CDGCN-based clustering method. First, we designed an initial speaker diarization system S6 with the Leiden clustering module only. The inputs of the system S6 are raw graphs, where every node pair has a weighted edge. Many node pairs are linked incorrectly, which causes the high Diarization Error Rate (DER) of the initial system.
Secondly, we applied the KNN algorithm to ensure that only the edges between the pivot node and its top-K neighbors are well-connected. This operation ignored many wrong linkages and made the DER decrease rapidly. The GCN refines the linkages between nodes according to their neighbors by adding GCN refinement within a sub-graph context. After refinement, the DER is decreased from 17.04\% to 15.38\% on the Full EVAL dataset. We performed the Graph-OSD module to further improve the system's performance, and achieved a DER of 13.72\% on the Full EVAL dataset. In order to evaluate the accuracy of second speaker labels produced by CDGCN, we used oracle OSD labels for the graph-OSD module. The results showed that the DER of the full EVAL dataset was improved from 13.72\% to 8.94\% significantly. This demonstrated that overlapped speech is a critical factor that limits system performance.
\begin{table}[]
\centering
\caption{MSE of speaker number prediction with different clustering methods on EVAL dataset.}
\label{tab:mse}
\begin{tabular}{llc}
\hline
ID & \multicolumn{1}{c}{Method} & MSE           \\ \hline
S1 & AHC                        & 3.80          \\
S2 & K-means                    & 2.05          \\
S3 & NME-SC                     & 2.05          \\
S6 & Raw-Leiden(ours)           & 4.45          \\
S7 & KNN-Leiden(ours)           & 2.38          \\
S5 & CDGCN(ours)                & \textbf{1.67} \\ \hline
\end{tabular}
\vspace{-0.5cm}
\end{table}

\vspace{-0.4cm}
\subsection{Speaker number prediction}
In order to further evaluate the performance of clustering methods, we calculated the Mean Square Error (MSE) of speaker number prediction for the above clustering methods. As shown in Table \ref{tab:mse}, the CDGCN outperformed the traditional clustering methods on the speaker number prediction task. The inputs of the KNN-Leiden system are speaker graphs constructed by KNN algorithm. When compared the performance of KNN-Leiden and CDGCN algorithms, we can see that GCN model boosts the MSE from 2.38 to 1.67 on EVAL dataset. By optimizing the global quality function, CDGCN can find a more appropriate graph partition to predict the number of speakers.

\section{Conclusions}
This paper proposes a novel speaker clustering method based on the speaker topological graph for speaker diarization. We aim to give consideration to both local and global information when clustering. The proposed CDGCN-based clustering approach include graph generation, sub-graphs detection, and Graph-OSD. The local linkage between speech segments is inferred by a GCN model, while the Leiden community detection algorithm is applied to find the global partition of the speaker graph.
To further improve the performance of our speaker diarization system, we also proposed a Graph-OSD component to handle overlapped speech for speaker diarization. Experimental results demonstrated that CDGCN based speaker diarization system outperformed conventional CSD systems in the DIHARD III corpus.

\label{sec:conclusions}

\bibliographystyle{IEEEbib}
\bibliography{strings,refs}

\end{document}